\documentclass[preprint,12pt]{elsarticle}
\journal{journal} \RequirePackage{graphicx}

\begin{document}

\begin{frontmatter}

\title{A generalized family of anisotropic compact object in general relativity}

\author{S.K. Maurya}
\address{Department of Mathematical and Physical Sciences,
College of Arts and Science, University of Nizwa, Nizwa, Sultanate
of Oman\\sunil@unizwa.edu.om}

\author{Saibal Ray}
\address{Department of Physics, Government College of Engineering \& Ceramic Technology, Kolkata 700010, West Bengal, India \\saibal@iucaa.ernet.in}

\author{Shounak Ghosh}
\address{Department of Physics, Indian Institute of Engineering Science and Technology, Shibpur, Howrah 711103, West Bengal,
India \\shnkghosh122@gmail.com}

\author{Sarbajit Manna}
\address{ Department of Materials Engineering,  Indian Institute of Science, Bangalore 560012, Karnataka, India,\\rick.sarba@gmail.com}

\author{Smitha T.T.}
\address{Department of Mathematical \& Physical Sciences, College of Arts \& Science, University of Nizwa, Nizwa, Sultanate of Oman \\smitha@unizwa.edu.om}

\date{Received: date / Accepted: date}

\maketitle

\begin{abstract}
We present model for anisotropic compact star under the general theory of relativity of Einstein. In the study a 4-dimensional spacetime has been considered which is embedded into the 5-dimensional flat metric so that the spherically symmetric metric has class 1. A set of solutions for the field equations are found depending on the index $n$ involved in the physical parameters. The interior solutions have been matched smoothly at the boundary of the spherical distribution to the exterior Schwarzschild solution which necessarily provides values of the unknown constants. We have chosen the values of $n$ as $n=2$ and $n$=10 to 20000 for which interesting and physically viable results can be found out. The investigations on the physical features of the model include several astrophysical issues, like (i) regularity behavior of stars at the centre, (ii) well behaved condition for velocity of sound, (iii) energy conditions, (iv) stabilty of the system via the following three techniques - adiabatic index, Herrera cracking concept and TOV equation, (v) total mass, effective mass and compactification factor and (vi) surface redshift. Specific numerical values of the compact star candidates LMC X-4 and SMC X-1 are calculated for central and surface densities as well as central pressure to compare the model value with actual observational data.
\end{abstract}


\end{frontmatter}

\section{Introduction}
Anisotropy factor is actually a function used in the Einstien field equation to consider the case of actual situation inside the star compared to the idealized isotropic case. According to Rago~\cite{Rago1991} the procedure to obtain the solution of field equations for anisotropic model, one can use two arbitary functions in general relativity and they are anisotropic factor ($\Delta=p_t-p_r$) and generating function. Pressure changes in this extreme conditions into two components - radial pressure and tangential pressure. From the expression of ansiotropy factor it is clear that the sign of it could be positive or negative and this has different significance, as such when the tangential pressure becomes greater than the radial pressure then the positive anisotropy factor generates an outwardly acting force. On the other hand, when the anisotropy factor is negative and hence the radial pressure is greater than tangential pressure, it generates an inward force.

Anisotropy comes to a system due to the inhomogeniety in pressures. There are lot of reasons for this pressure anisotropy, however the main reasons are very high density region in the core region, various condensate states (like pion condensates, meson condensates etc.), superfluid 3A, mixture of fluids of different types, rotational motion, presence of magnetic field and phase transition etc. According to Bowlers and Liang~\cite{BL1974} there is no celestial object completely made of perfect fluid and specially  in the system like compact stars, enormous density and gravitational pull give birth to even larger amount of anisotropy than other normal stars. The idea of this anisotropic fluid pressure was originally proposed by Ruderman~\cite{Ruderman1972} and later on by Canuto~\cite{Canuto1973}. However, Herrera and Santos~\cite{HS1997}  in a review paper have discussed various aspects of the anisotropy. For further study related to pressure anisotropy on diversified topics several other works are available in the literature~\cite{Ivanov2002,SM2003,MH2003,Usov2004,Varela2010,Rahaman2010,Rahaman2011,Rahaman2012,Kalam2012,Deb2015,Shee2016,Maurya2016,Maurya2017}.  

It is also to note that the chances of having anisotropy is much higher in compact star because the interaction among the particles is too relativistic and due to this relativistic motions of the particles, they become too random to main any uniform distribution throughout the region and thus this relativistic nature of particles in compact star could be one of the possible reason for giving birth of significant anisotropy in the compact star. Actually anisotropy generates an anisotropic force and the anisotropy throughout the system effects the density inside the star. So the solution becomes an anisotropic sphere with variable density and we will discuss the nature of this density for anisotropic compact star later. This anisotropic force as we said is positive for the outward force can also be called repulsive in nature and this repulsive nature of anisotropic force inside the star makes the compact object more compact than the isotropic condition. So anisotropy is one of the reason for making the compact star more compact.

 In the present paper to study the above mentioned compact star a class 1 condition for spherically symmetric metric have been considered. This line element will provide us a primary tool for solving the nonlinear Einstein field equations. In connection to the metric we would like to mention here that according to Eddigton~\cite{Eddigton1924} the 4-dimensional surfaces in higher dimension will not change the metric at all. In the present approach therefore we have used the same concept here for getting the expression of metric coefficients by considering a 5-dimensional flat spacetime metric. We consider here the manifold of the spacetime curve in a flat Euclidean 5-dimensional metric and  proceed with the same spherical polar coordinate to reach at the expression for the coefficients. In our model, we have used class 1 condition for solving the field equations which basically tells us that 4-dimensional spacetime can be embeeded in 5-dimensional pseudo-Euclidean space. We follow the Karmarker condition~\cite{Karmarkar1948} for using the class 1 metric and the relationship between $\nu$ and $\lambda$ has been achieved. It can be seen that this will generate new solutions of compact stars of embedding class 1 spacetime. To get a suitable mathematical exposition on the $n$ dimensional manifold which can be embedded in a pseudo-Euclidean space of $m = n(n + 1)/2$ dimensions is available in the following works and Refs. therein~\cite{Schlaefli1871,Kuzeev1980,SKM1,SKM2}. The main point of this whole discussion is that the choice of coordinate system is always a major issue in general relativistic systems. The inside of the compact object is obyed by interior the Schwarzschild metric whereas beyond the surface of the object where radial pressure is zero, the exterior Schwarzschild metric is applicable though this is not our concern as we shall only deal with compact star. 

The plan of our investigation is as follows: In the Sect. 2 we have provided basic formalism of (i) class 1 metric, and (ii) Einstein's field equations in connection to anisotric fluid sphere. Sect. 3 deals with the solution of Einstein's field equations for different physical parameters, viz. the gravitational potentials, fluid pressure and energy density. We have explored and discussed several physical features in Sec. 4 and a comparative study has been conducted in Table 1 and 2. In the last Sect. 5 we have drawn some concluding remarks with some salient features of the present model.

\section{Basic mathematical formalism}

\subsection{Class 1 condition}
Let us consider the line element of the spherically symmetric metric for a star as
\begin{equation}
ds^{2}=-e^{\lambda(r)}dr^{2}-r^{2}\left(d\theta^{2}+\sin^{2}\theta d\phi^{2} \right)+e^{\nu(r)}dt^{2}, \label{eq1}
\end{equation}
where $\lambda$ and $\nu$ are the functions of the radial coordinate $r$. 

We now suppose the 5-dimensional flat metric in the form
\begin{equation}
ds^{2}=-\left(dz^1\right)^2-\left(dz^2\right)^2-\left(dz^3\right)^2-\left(dz^4\right)^2+\left(dz^5\right)^2,\label{eq2}
\end{equation}
where we have supposed that $z^1=r\,sin\theta\,cos\phi$, $z^2=r\,sin\theta\,sin\phi$, $z^3=r\,cos\theta$, $z^4=\sqrt{4C}\,e^{\frac{\nu}{2}}\,cosh{\frac{t}{\sqrt{4C}}}$, $z^5=\sqrt{4C}\,e^{\frac{\nu}{2}}\,sinh{\frac{t}{\sqrt{4C}}}$ with $C$ as a constant.

The differential of the above components are expressed as follows:
 \begin{equation}
dz^1=dr\,sin\theta\,cos\phi + r\,cos\theta\,cos\phi\,d\theta\,-r\,sin\theta\,sin\phi\,d\phi,\label{eq3}
\end{equation}

\begin{equation}
dz^2=dr\,sin\theta\,sin\phi + r\,cos\theta\,sin\phi\,d\theta\,+r\,sin\theta\,cos\phi\,d\phi,\label{eq4}
\end{equation}

\begin{equation}
dz^3=dr\,cos\theta\, - r\,sin\theta\,d\theta,\label{eq5}
\end{equation}

\begin{equation}
dz^4=\sqrt{4C}\,e^{\frac{\nu}{2}}\,\frac{\nu'}{2}\,cosh{\frac{t}{\sqrt{4C}}}\,dr + e^{\frac{\nu}{2}}\,sinh{\frac{t}{\sqrt{4C}}}\,dt,\label{eq6}
\end{equation}

\begin{equation}
dz^5=\sqrt{4C}\,e^{\frac{\nu}{2}}\,\frac{\nu'}{2}\,sinh{\frac{t}{\sqrt{4C}}}\,dr + e^{\frac{\nu}{2}}\,cosh{\frac{t}{\sqrt{4C}}}\,dt.\label{eq7}
\end{equation}

By substuting the values of the differential components from Eqs. (\ref{eq3}) to (\ref{eq7}) into the metric (\ref{eq2}), we get
\begin{equation}
ds^{2}=-\left(\,1+C\,e^{\nu}\,{\nu'}^2\,\right)\,dr^{2}-r^{2}\left(d\theta^{2}+\sin^{2}\theta d\phi^{2} \right)+e^{\nu(r)}dt^{2}.\label{eq8}
\end{equation}

The metric (\ref{eq3}) represents the metric (\ref{eq1}) under the condition
\begin{equation}
e^{\lambda}=\left(\,1+C\,e^{\nu} \,{\nu'}^2\,\right).\label{eq9}
\end{equation}

The Eq. (\ref{eq8}) implies that we can embed	a 4-dimensional spacetime into the 5-dimensional flat metric. Then our spherical symmetric metric (\ref{eq1}) has class 1 when the condition (\ref{eq9}) is satisfied.

By solving Eq. (\ref{eq9}), we get the value of $e^{\nu}$ as
\begin{equation}
e^{\nu}=\left(\,A+\frac{1}{\sqrt{4C}}\,\int{\sqrt{e^{\lambda}-1}}\,\right)^2,\label{eq10}
\end{equation}
where $A$ is a constant of integration.

The metric coefficients $e^\lambda$ and $e^\nu$, as expressed in Eqs. (\ref{eq9}) and (\ref{eq10}), have much significance as the value of these two coefficients helps us to realize we are in which law. As our compact object is spherically symmetric, we have used the line element that is given in Eq. (\ref{eq1}) being the most general centrally symmetric expression for $ds^2$. Here $g_{00}=e^\nu$ and $g_{11}=-e^\lambda$ such that only the magnitudes of these metric coefficients can change the scenario completely.

\subsection{Einstein's field equations}
To describe Einstein's field equations of general relativity once Wheeler~\cite{Wheeler1998} stated that ``Spacetime tells matter how to move, matter tells spacetime how to curve.''  The matter-energy density determines how much the curvature of spacetime will be and plays an important role for determining the matter distribution inside the compact star as it contains density of energy and momentum in spacetime. In compact star (which is a highly gravitating object and needs general relativity to study) the stress-energy tensor is a source of spacetime curvature. So in general, $T_{ij}$ determines the degree of curvature and inside the star can be written in the following standard form 
\begin{equation}
T_{ij}=diag(\rho,-p_r,-p_t,-p_t),\label{eq11}
\end{equation}
where $\rho$ is the density, $p_r$ and $p_t$ are the radial and tangential pressures respectively. Note that the stress-energy tensor is always a diagonal matrix where the first component is relativistic tensor term which is actually the matter-energy density because at very high speed or in relativistic condition, $v\approx c$, so the expression of it becomes energy (i.e. $T_{00}=\rho c^2$, considering $c=1$, it will basically be the density term, so that $T_{00}=\rho$) and the other three diagonal terms are pressures. For isotropic fluid case, all these pressures are the same, as there is no anisotropy. But in the present case where we are interested for studying the anisotropy in the system, second component is the radial pressure and last two components are the tangential pressures. 

Therefore the Einstein field equations for the anisotropic field distributions can be provided as 
\begin{equation}
R_{ij}-\frac{1}{2}Rg_{ij}=-8 \pi T_{ij}, \label{eq12}
\end{equation}
where $R_{ij}$ are the Ricci tensor and $R$ is the Ricci saclar of the curvature.

With the above specifications the set of Einstein field equations for the metric~(1) can be represented explicitly as follows
\begin{equation}
\frac{1-e^{-\lambda}}{r^{2}}+\frac{e^{-\lambda}\lambda'}{r}=8\pi\rho,\label{eq13}	
\end{equation}

\begin{equation}
\frac{e^{-\lambda}-1}{r^{2}}+\frac{e^{-\lambda}\nu'}{r}=8\pi\, p_{r}, \label{eq14}
\end{equation}

\begin{equation}	
e^{-\lambda}\left(\frac{\nu''}{2}+\frac{\nu'^{2}}{4}-\frac{\nu'\lambda'}{4}+\frac{\nu'-\lambda'}{2r} \right)=8\pi\, p_t.\label{eq15}
\end{equation}

From the above field equatins (\ref{eq13} - \ref{eq15}) it is obvious that $\rho$ and $p$ vary throughout the spacetime of compact star and so they act as ingredient of field.

\section{Generating new class of anisotropic solutions}
 We have supposed the metric coefficient $e^{\lambda}$ to obtain the general class of anisotropic solution, which is given as
\begin{equation}
e^{\lambda}=1+\frac{ar^2}{(1-br^2)^{2n}},\label{eq16}
\end{equation}
where $a$ and $b$ are the parameters with units of $length^{-2}$. It is to note that Singh and Pant~\cite{SP2016} have used similar type of metric coefficient for positive value of $n$, however our choice of $n$ here is negative as the factor is coming in the denominator of the fraction. Now from Eq. (\ref{eq16}), at $r \rightarrow 0$, we have $e^\lambda=1$ which implies that the metric potential $\lambda$ is physically valid as it is free from singularity at the centre of the physical system.

However, the value of $e^{\nu}$ is determined by using the Eqs. (\ref{eq16}) and (\ref{eq10}) as
\begin{equation}
e^{\nu}=[A+B\,(1-br^2)^{1-n}]^2,\label{eq17}
\end{equation}
where $B=-\frac{\sqrt{a}}{4\,b\,\sqrt{C}\,(1-n)}$ and $n > 1$. 

Now from Eq. (\ref{eq17}), it is simple to observe that at $r \rightarrow \infty$, we have $e^\nu=A^2$ where $A\neq0$. Now if $A=1$, both of the metric coefficients will be unity, i.e. at very large distance from the gravitating fields (or bodies) the metric is no longer behaving like the Schwarzschild metric as we use it for applying general relativity but it behaves more (or it becomes) like the Galilean metric criteria. In this ``far from the field'' situation, we do not have to worry anything about general relativity. Here as the situation is for large distances we can always very correctly assume the gravitation field is weak and we will have $g_{00}=1+2\phi/c^2$ (where $\phi$ is the Newtonian gravitational potential and $c$ is the speed of light) and for this situation we can make the second term as zero hence $g_{00}$ will be unity and we can relate it with our previous discussion in terms of metric coefficient.

The expression of energy density is obtained by plugging the value of $e^{\lambda}$ into the Eq. (\ref{eq13}), which is given as
\begin{equation}
{8\pi\,\rho}=\frac{a\left[ar^2\,(1-br^2)+3(1-br^2)^{2n+1}+4nbr^2\,(1-br^2)^{2n}\right]}{(1-br^2)\,[ar^2+(1-br^2)^{2n}]^2}. \label{eq18}
\end{equation}

Again, the radial pressure ($p_r$) and tangential pressure ($p_t$) are determined by using Eqs. (\ref{eq16}) and (\ref{eq17}) into Eqs. (\ref{eq14}) and (\ref{eq15}) as
\begin{equation}
{8\pi\,p_r}=\frac{4bB(n-1)(1-br^2)^{2n}-a[B(1-br^2)+A(1-br^2)^n]}{[B(1-br^2)+A(1-br^2)^n][ar^2+(1-br^2)^{2n}]}, \label{eq19}
\end{equation}

\begin{equation}
{8\pi\,p_t}=\frac{(1-br^2)^{2n-1}\left[4bB\,(n-1)\,p_{t1}+a\,p_{t2}+a\,B(b^2r^4-1)\right]}{[B(1-br^2)+A\,(1-br^2)^n]\,[ar^2+(1-br^2)^{2n}]^2}, \label{eq20}
\end{equation}
where $p_{t1}=(1-br^2)^{2n}(1-br^2+nbr^2)$, $p_{t2}=-A(1-br^2)^n(1-br^2+2nbr^2)$.

As there is no singularity, the metric and the spacetime itself is well defined even at the centre of the object, i.e. we can use the coordinate systems to define its poperties. We can say also that at the centre though the density and pressures are very high (as is evident from Eqs.(\ref{eq18}) -(\ref{eq20})) the gravitating force not becomes so high that it completely curves the spacetime into singularity (in contrast with black hole). 

By subtracting Eq. (\ref{eq19}) from Eq. (\ref{eq20}), the expression for anisotropy factor, $\Delta=p_t - p_r$, is given as
\begin{equation}
\Delta= \frac{r^2\,\left[2bn(1-br^2)^{2n-1}-a\right]\left[2bB(n-1)(1-br^2)^{2n}-aB(1-br^2)-aA(1-br^2)^{n}\right]}{8\,\pi\,\left[ar^2+(1-br^2)^{2n}\right]^2\,\left[B(1-br^2)+A\,(1-br^2)^n\right]}. \label{eq21}
\end{equation}

From the above expression it is obvious that the anisotropy factor depends not only on the factors as mentioned in the introductory section  but also on the constants $A$, $B$, $n$, $a$, $b$ and of course on the radius $r$. So for different compact stars, the anisotropy factor differs largely. From the expression of ansiotropy factor it is clear that the sign of it could be positive or negative and this has different significance. Positive anisotropy factor, i.e. when the tangential pressure is greater than the radial pressure, the anisotropy factor generates an outward force whereas when the anisotropy factor is negative, i.e. when radial pressure is greater than the tangential pressure, it generates an inward force. 

From Eq. (\ref{eq21}), it is clear that at the centre, i.e. at $r=0$, the anisotropy factor vanishes. It is indicating then that the radial and tangential pressures are same at the centre of the compact star. Anisotropy generates an anisotropic force and the anisotropy throughout the system effects the density inside the star. So the solution becomes an anisotropic sphere with variable density and we shall discuss the nature of this density for anisotropic compact star later on. This anisotropic force as we said is positive for the outward force can also be called repulsive in nature and this repulsive nature of anisotropic force inside the star makes the compact object more compact than the isotropic condition. So anisotropy is one of the reason for making the compact star more compact. This can be explained like at very high core density, the particles condense together and they love this condense state and as they don't want to move apart from each other, radial pressure or force in the radial direction becomes insignificant than the force along the tangential direction and this means the compactness of the compact star increases which actually a result of positive anisotropy inside the star (mainly in the core region). It could be claimed from this point that during the birth of compact star when they are little bit hotter, the abscence of condensate states may increase the radial pressure which makes the negative anisotropy factor and hence lower compactness. So with time the tendency of the positive anisotropic factor increases inside the star.

\section{Matching condition}
Exterior spacetime describes the spacetime outside a spherically symmetric object which, in the neutral case, can be defined by the Schwarzschild metric as given in the expression below
\begin{equation}
ds^{2} =-\left(1-\frac{2M}{r} \right)^{-1} dr^{2}-r^{2} (d\theta ^{2} +\sin ^{2} \theta \, d\phi ^{2} )+ \left(1-\frac{2M}{r} \right)\, dt^{2},\label{22}
\end{equation}
where $M$ is the total mass of the star.

If we observe very carefully, then it will reveal that this Schwarzschild metric is very simillar with the spherically symmetric line element metric that we have expressed in Eq. (\ref{eq1}). This means that this is also spherically symmetric metric but with different metric coefficients. The following relations by comparing the metric coefficients: $e^{-\lambda}=(1-\frac{2 M}{r})$ and $e^\nu=(1-\frac{2 M}{r})$. The Schwarzschild radius is defined by $R_s=\frac{2 G M}{c^2}$ or $2M$ (as $G=c=1$ in geometrized unit). If $R_s<r$ always then the singularity will be outside the spherically symmetric domain and this is very much physically acceptable as we already know that there could not be any singularity inside in this system. So the solutions will be non-singular throughout the region of the domain. In corresponding to the interior spacetime we therefore employed the exterior spacetime described by the Schwarzschild metric.

The radial pressure must be zero at the surface of the compact star ($p_r~(r=R)=0$). The reason could be that the surface is the boundary of a gravitating object and there is no further extension of that gravitating object (or particles of that object) beyond the bounding surface and hence at the surface there is no more radial pressure which can be directed outwards. So for the vanishing radial pressure at the surface~\cite{Misner1964} we have
\begin{equation}
\frac{A}{B}=\frac{4b\,(n-1)\,(1-bR^2)^{2n}-a\,(1-bR^2)}{a\,(1-bR^2)^n}. \label{eq23}
\end{equation}

Again at the boundary of the star (i.e. $r=R$) we have the condition $e^{\nu(R)}=e^{-\lambda(R)}$. So using this condition and 
Eq. (\ref{eq23}) we obtain the value of the constant $B$ as
\begin{equation}
B=\frac{a}{4b(n-1)\sqrt{aR^2+(1-bR^2)^{2n}}}. \label{eq24}
\end{equation}

Now by putting the value of $B$ into Eq. (\ref{eq23}), we get
 \begin{equation}
A=\frac{4b\,(n-1)\,(1-bR^2)^{2n}-a\,(1-bR^2)}{4\,b\,(1-bR^2)^n\,\sqrt{aR^2+(1-bR^2)^{2n}}}. \label{eq25}
\end{equation}

From the expression of Eq. (\ref{eq23}), it is clear that the constants $A$ and $B$ are functions of $a$, $b$, $n$ and $r$, i.e. they are dependent on the same parameters like $p_r$, $p_t$ and $\rho$.

\section{Physical features of the models}

\subsection{Regularity behavior of stars at the centre}

(i) As we have mentioned earlier that by putting $r=0$, i.e. at the centre of the star in Eqs. (\ref{eq16}) and (\ref{eq17}), we can have $e^{\nu(0)}=(A+B)^2$ and $e^{\lambda(0)}=1$. These values of metric coefficients of anisotropic compact star are physically acceptable as they are finite at the centre without having any singularity. So the anisotropic compact star will be stable at this point. This is one of the main difference between compact star and black hole where black hole achieves singularity at the centre so that the metric coefficients become infinite for black hole.

%%%%%%%%%%%%%%%%%%%%%%%%%%%%%%%%%%%%%%%%%%%%%%%%%%%%%%%%%%%%%%%
\begin{figure}[!h]\centering
    \includegraphics[width=6.5cm]{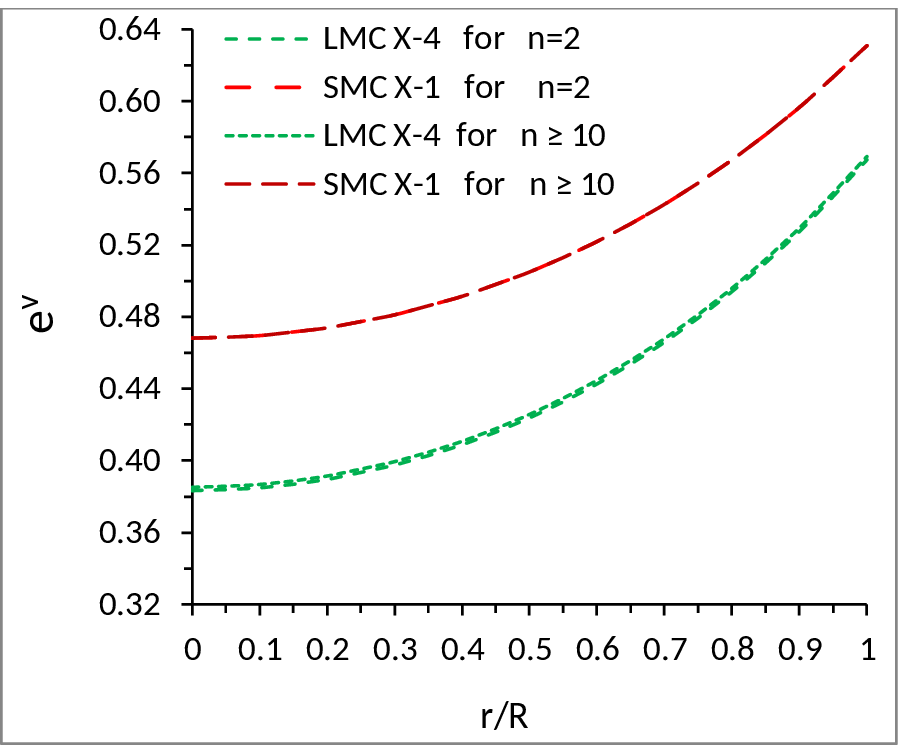} \includegraphics[width=6.5cm]{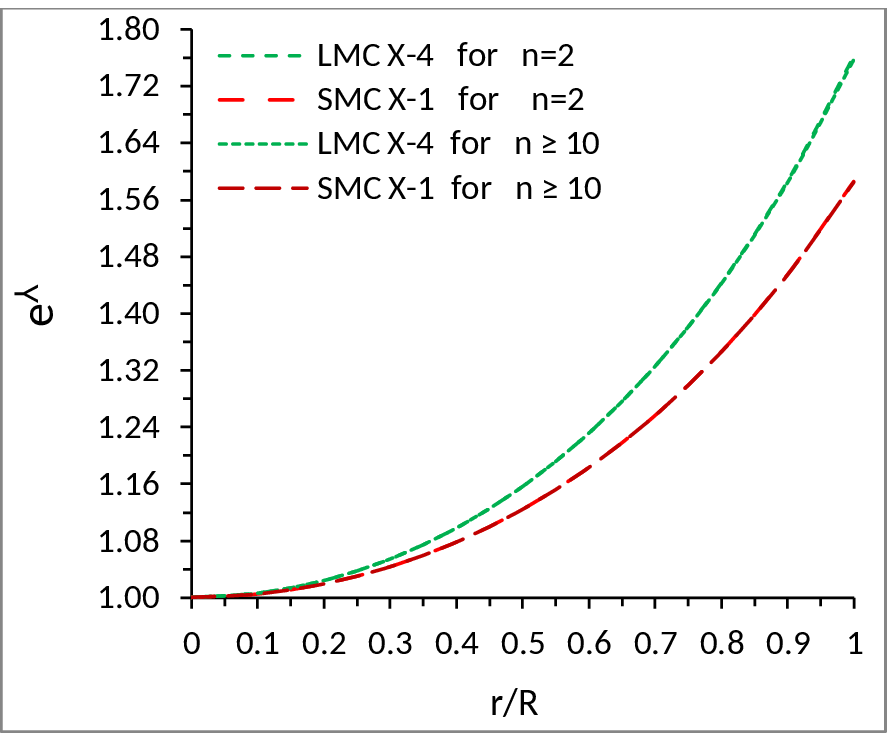}
\caption{The Behavior of metric function $e^{\nu}$ (left panel) and $e^{\lambda}$ (right panel) with the fractional coordinate $r/R$. The numerical values of the parameters and arbitrary constants for different compact stars are as follows: (i) LMC X-4 : $a=0.0075$, $b=0.000821$ for $n=2$ and $a=0.0075$, $nb=0.00164$ for $n\ge 10$, (ii) SMC X-1: $a=0.00681$, $b=0.00078$ for $n=2$, and $a=0.00681$, $nb=0.00159$ for $n \ge 10$. The same specifications will be followed in all the later plots. }
    \label{metricfunction}
\end{figure}
%%%%%%%%%%%%%%%%%%%%%%%%%%%%%%%%%%%%%%%%%%%%%%%%%%%%%%%%%%%%%%%

(ii) Pressure and density both are very high for compact star comparing with the other stars. In our anisotropic compact star pressure is not throughout uniform and there are two component pressure in the radial as well as tangential directions. It can be observed from the  expressions of our model that the radial and tangential pressures both are the same in magnitude at the centre of the anisotropic compact star, i.e. there is no anisotropy at the centre ($\Delta=0$) of the compact star. The expressions of $p_r$ and $p_t$ make it clear that they are independent of radius at the centre of our anisotropic compact star. This means that$\frac{dp_r}{dr}=\frac{dp_t}{dr}=0$ at the centre and also $\frac{d^2p_r}{dr^2}<0,\frac{d^2p_t}{dr^2}<0$. These relations give the conclusion that both radial and tangential pressure reach the maximum value at the centre of the anisotropic compact star.

Now the radial pressure and tangential pressure at the centre are obtained from Eqs. (19)  and (20) as follows:
 \begin{equation}
 p_r(0)=\frac{4bB(n-1)-a(A+B)}{8\pi (A+B)},\nonumber\\
\end{equation}

\begin{equation}
  p_t(0)=\frac{4bB(n-1)-a(A+B)}{8\pi (A+B)}.\nonumber\\
\end{equation}

For any physically acceptable model, $p_r$ and $p_t$ must be positive and finite at the centre. This condition gives
\begin{equation}
\frac{A}{B} < \frac{4\,b\,(n-1)-a}{a}.  \label{eq27}
\end{equation}

%%%%%%%%%%%%%%%%%%%%%%%%%%%%%%%%%%%%%%%%%%%%%%%%%%%%%%%%%%%%%%%
\begin{figure}[!h]\centering
    \includegraphics[width=6.5cm]{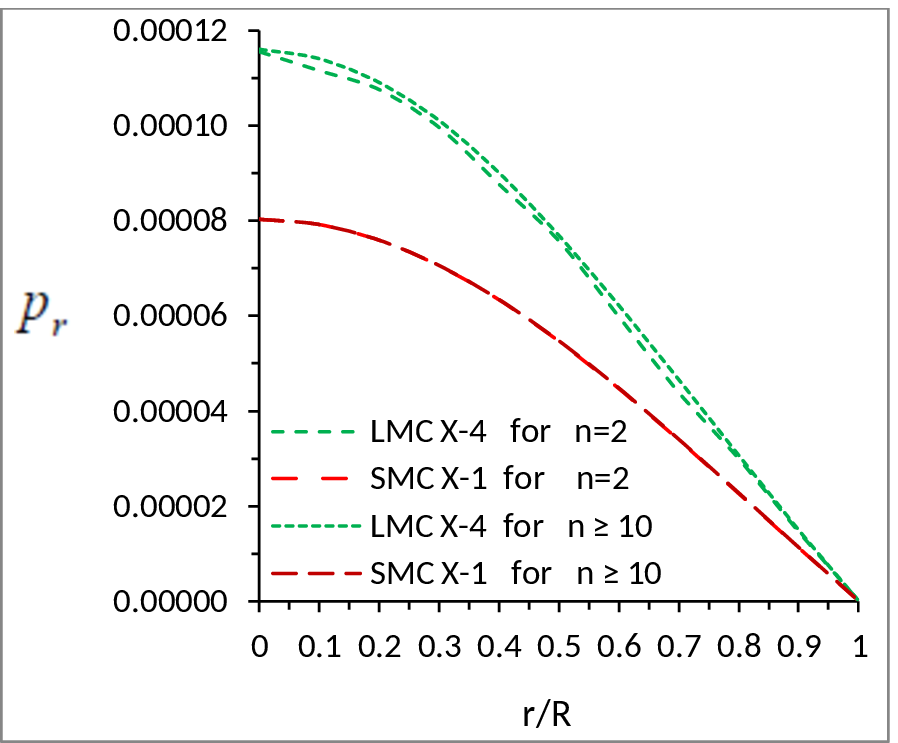} \includegraphics[width=6.5cm]{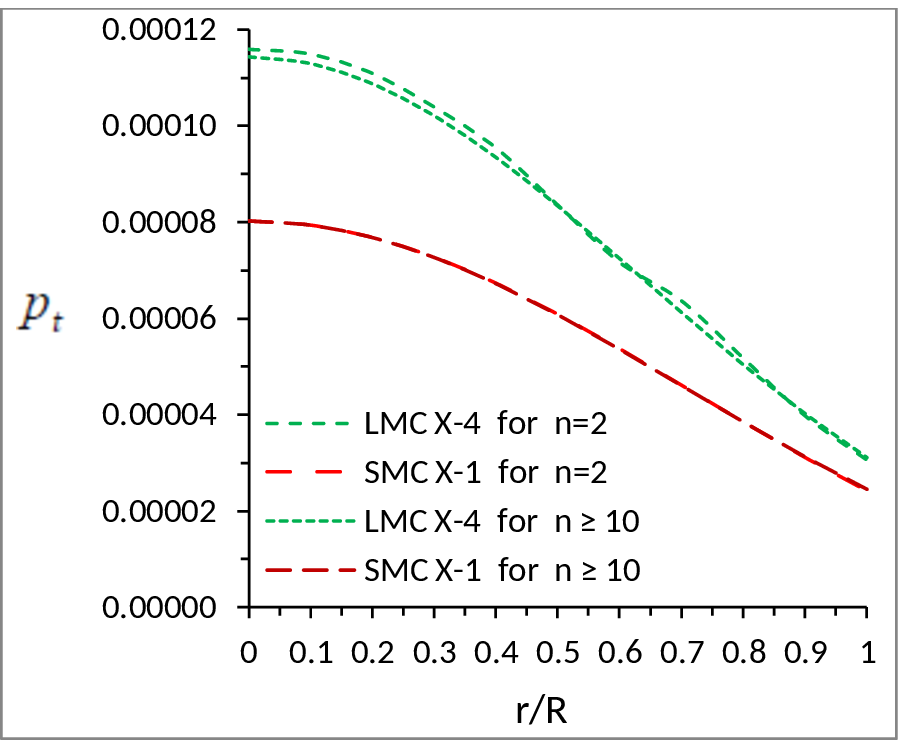}
\caption{The Behavior of radial pressure $p_r$ (left panel) and tangential pressure $p_t$ (right panel) with the fractional coordinate $r/R$. }
    \label{pressure}
\end{figure}
%%%%%%%%%%%%%%%%%%%%%%%%%%%%%%%%%%%%%%%%%%%%%%%%%%%%%%%%%%%%%%%

(iii) Density behaves in the same way like pressure for anisotropic compact star i.e. it reaches the maximum at the centre and then decreases smoothly upto the surface. From Eq. (18) matter density at centre is  $\rho(0)=\frac{3\,a}{8\,\pi}>0$, that implies $a$ must be positive and vice versa. Now $a$ being positive in magnitude the central pressure is also positive and finite which is very much physically acceptable because for making the compact star stable against gravitational collapse the central pressure must be positive. 

The variation of the density with the fractional coordinate is shown in Fig. 3. The density is positive throughout the anisotropic compact star in our model as it has been shown in this figure but it gradually decreases from centre to surface.In the figure we can see that density is constant upto certain distance from the centre. One reason behind this feature could be the assymtotic freedom of quraks due to MIT bag model. The quarks could not free themseleves beyond  a certain distance from the centre and this causes almost a constant density in that region.

%%%%%%%%%%%%%%%%%%%%%%%%%%%%%%%%%%%%%%%%%%%%%%%%%%%%%%%%%%%%%%%
\begin{figure}[!h]\centering
    \includegraphics[width=6.5cm]{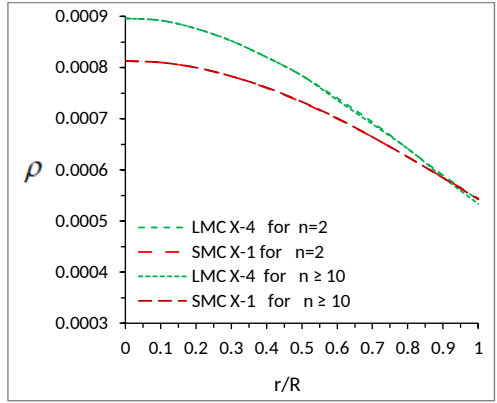}
\caption{The Behavior of energy density $\rho$ with the fractional coordinate $r/R$. }
    \label{density}
\end{figure}
%%%%%%%%%%%%%%%%%%%%%%%%%%%%%%%%%%%%%%%%%%%%%%%%%%%%%%%%%%%%%%%

(iv) The solution of anisotropic compact star must satisfy $p_r/\rho <1$ and $p_t/\rho < 1$ which are known as the Zel'dovich condition. It means that higher the density of a compact star, more will be the magnitude of the radial and tangential pressure but magnitude of these pressures could not go beyond the magnitude of density of the compact star. Here density means the total number of particles in that certain volume where this individual pressure arises from some part of those particles and not all. Hence with the increase of density radial and tangential pressures increase but the magnitude of them must be lower than the density. 

By following the above mentioned Zeldovich's condition we derived the inequalities, from Eqs. (18) - (20), as follows 
\begin{equation}
  \frac{b(n-1)-a}{a} < \frac{A}{B}. \label{eq28}
\end{equation}

Using the Eqs. (\ref{eq27}) and (\ref{eq28}), we again get
\begin{equation}
  \frac{b\,(n-1)-a}{a} < \frac{A}{B} < \frac{4\,b\,(n-1)-a}{a},  \label{eq29}
\end{equation}
where $n>1$.

In connection to Eq. (\ref{eq27}) we desribed earlier that lower the value of $B$, more stable will be the system. However, the situation now is much more critical that both $A$ and $B$ have to follow the inequality as is expressed in (\ref{eq29}). So all the parameters, viz. $a$, $b$, $A$, $B$, $n$, can not reach below a minimum or beyond a maximum value to make a stable anisotropic compact star.

\subsection{Well behaved condition for velocity of sound}
It is expected that in any anisotropic compact star both the radial and tangential sound velocity are maximum at the centre and monotonically decrease upto the surface. Here the centre is isotropic and hence pressure is uniform, it does not divide into two parts which means the magnitude of pressure is very high and the differential pressure is very large in magnitude at the centre which creates very high velocity of sound. In compact star  the velocity of sound in radial and tangential directions decreases away from the centre because the interaction energy among the particles decreases. 

The radial velocity $(V_r)$ and the tangential velocity $(V_t)$ of the sound can be obtained respectively by the variation of radial and tangential pressures with density, i.e.
\begin{eqnarray}
V^2_r&=&\frac{(n-1)\,B\,F^2(r)\,f^2\,\left[8b^2\,G(r)\,\,n\,f^{2n-1}+4b^2\,H(r)\,\,f^{2n}\right]-F(r)\,f^2\,I_{1}(r)}
{a\,G^2(r)\left[a^2\,f^2\,r^2-2\,b\,n\,f^{4n}(5-3\,b\,r^2+4\,n\,b\,r^2)+I_{3}(r)\right]}\\ \label{eq30}
V^2_t&=&\frac{f^{2n}\,\left[\,L_{1}(r)\,L_{2}(r)+b\,G(r)\,F(r)\,L_{3}(r)+b\,G(r)\,F(r)\,L_{4}(r)\,\right]}
{a\,G^2(r)\left[\,a^2\,f^2\,r^2-2\,b\,n\,f^{4n}(5-3\,b\,r^2+4\,n\,b\,r^2)+I_{3}(r)\,\right]}, \label{eq31}
\end{eqnarray}
where\\
     $f=(1-br^2)$,\,\,\, $F(r)=\left(ar^2+f^{2n}\right)$, \,\,\,$G(r)=\left(B\,f+A\,f^n\right)$ ,\\
     $I_{1}(r)=[G(r)\,\left(a-2bnf^{2n-1}\right)\,I_{2}(r)]$, \,$I_{2}(r)=\left[-4bB(n-1)\,f^{2n}+a\,G(r)\right]$,\\
     $H(r)=-\left(B+A\,n\,f^{n-1}\right)$,\,  $I_{3}(r)=a\,f^{2n}\,[5+2b\,(3n-5)r^2+(5-10n+8n^2)b^2r^4]$, \\
     $L_{1}(r)=\left[\,4\,b\,B\,(n-1)\,f^{2n}\,(f+n\,b\,r^2)-a\,A\,f^n\,(f+2\,n\,b\,r^2)+a\,B\,(-1+b^2\,r^4)\,\right]$, \\
     $L_{2}(r)=\left[\,b\,f\,H(r)\,F(r)+b\,(2n-1)\,G(r)\,F(r)+2\,f\,G(r)\,\left(a-2\,b\,n\,f^{2n-1}\right)\,\right]$,\\
    $L_{3}(r)=\left[\,4\,b\,B\,(n-1)\,f^{2n}\,(f+n-b\,n\,r^2+2\,b\,n^2\,r^2)+2\,a\,B\,b^2\,r^4 \,\right]$, \\
    $L_{4}(r)=\left[a\,A\,(n-1)\,f^n-2\,a\,B\,b\,r^2-A\,a\,b\,r^2\,(2n^2+n-1)\,f^n \right]$.\\

Now the causality conditions demands that the radial and transverse velocity of sound must be less than unity. The variation of  $V_r$ and  $V_t$ with the fractional coordinate $r/R$ are shown in Fig. 4. It can be observed from this figure that the values of $V_r$ and  $V_t$ are lies between 0 and 1 everywhere inside the stellar system. Fig. 4 also reveals that both the radial and tangential pressures are non zero at the surface of the anisotropic compact star as well as the radial velocity of sound is always larger than the tangential velocity of sound.

%%%%%%%%%%%%%%%%%%%%%%%%%%%%%%%%%%%%%%%%%%%%%%%%%%%%%%%%%%%%%%%
    \begin{figure}[!h]\centering
    \includegraphics[width=6.5cm]{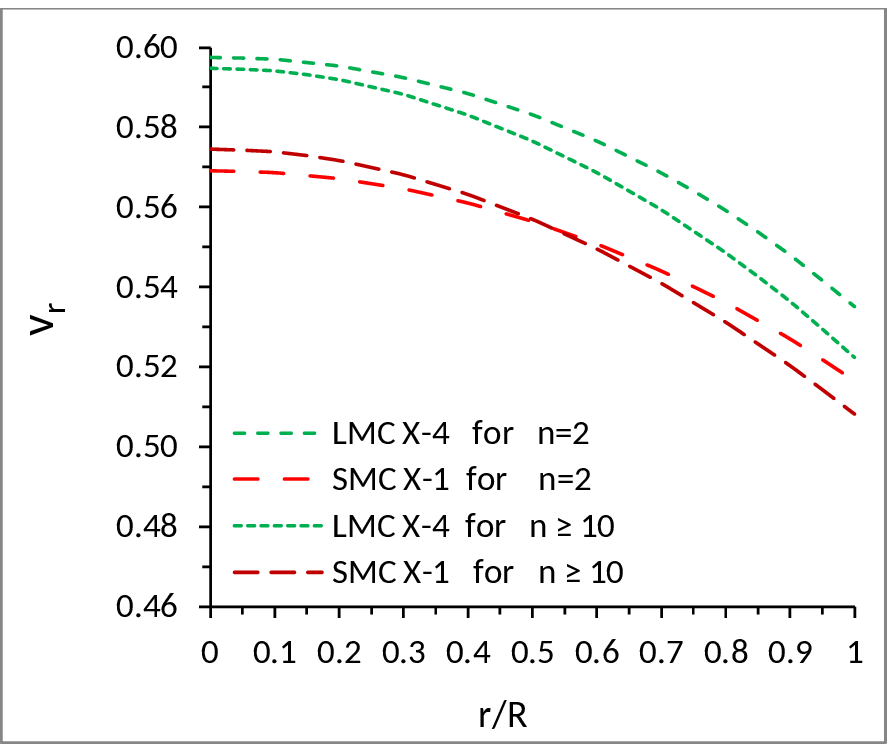} \includegraphics[width=6.5cm]{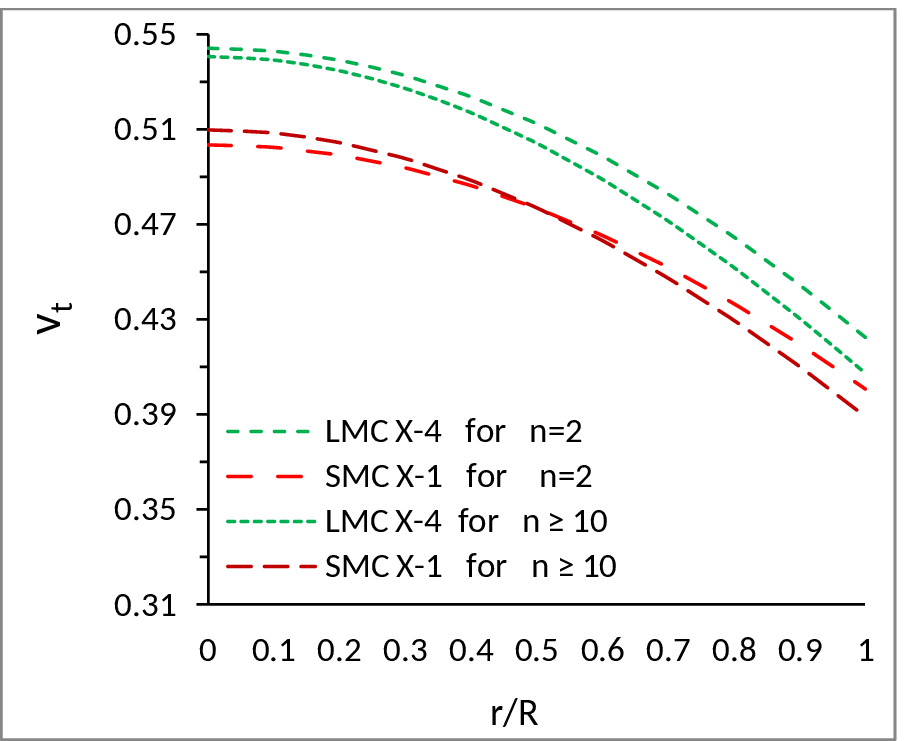}
\caption{The Behavior of radial velocity of sound $V_r$ (left panel) and tangential velocity of sound $V_t$ (right panel) with the fractional coordinate $r/R$. }
    \label{velocity}
\end{figure}
%%%%%%%%%%%%%%%%%%%%%%%%%%%%%%%%%%%%%%%%%%%%%%%%%%%%%%%%%%%%%%%

 \subsection{Energy conditions}
For any physically acceptable anisotropic fluid sphere the energy conditions, viz. Null Energy Condition (NEC), Strong Energy Condition (SEC) and Weak Energy Condition (WEC) are to be satisfied by the matter inside it. The actual meaning of all these energy conditions is that energy can never be negative as negative energy condition could never make stable condition inside the system. These conditions are hold if the following inequalities are satisfied:
\begin{equation}
(i)\,\,NEC: \rho\geq 0,\label{eq32}
\end{equation}

\begin{equation}
(ii)\,\,SEC: \rho-p_r-2{{p}_{t}} \geq  0, \label{eq33}
\end{equation}

\begin{eqnarray}
(iii)\,\,WEC: \rho- {{p}_{r}} \geq  0 \hspace{0.1cm}({WEC}_{r}) \hspace{0.2cm} and
     \hspace{0.2cm} \rho - p_t \geq  0 \hspace{0.1cm}({WEC}_{t}). \label{eq34}
\end{eqnarray}

 The variations of different energy conditions with the fractional parameter
 are shown in Fig. 5 where from all the plots it has been observed that energy
 conditions are satisfied in our model.

%%%%%%%%%%%%%%%%%%%%%%%%%%%%%%%%%%%%%%%%%%%%%%%%%%%%%%%%%%%%%%%
\begin{figure}[!h]\centering
    \includegraphics[width=6.5cm]{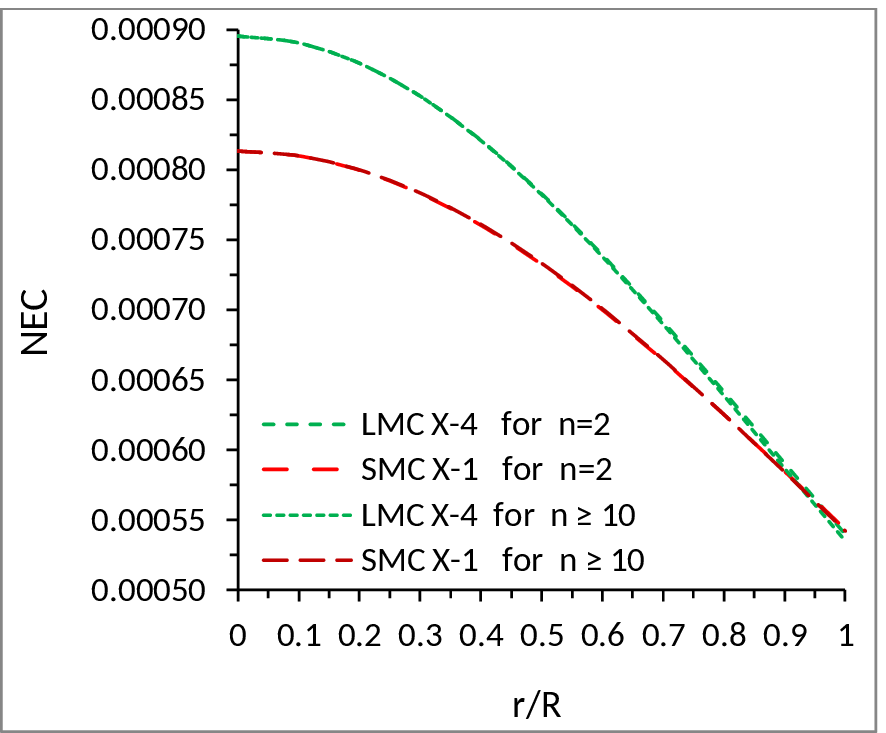} \includegraphics[width=6.5cm]{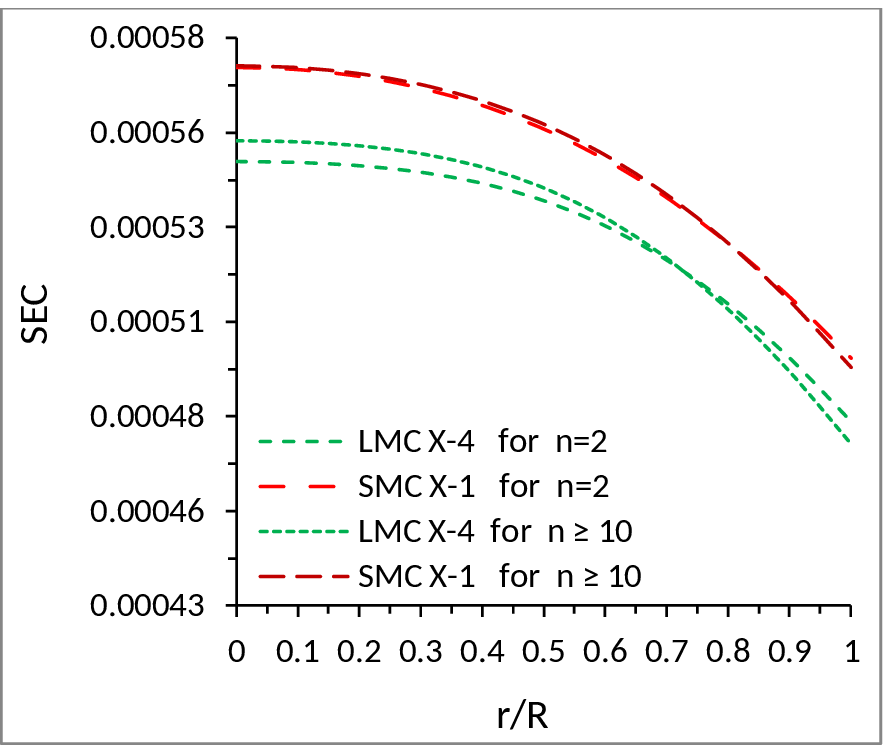}
    \includegraphics[width=6.5cm]{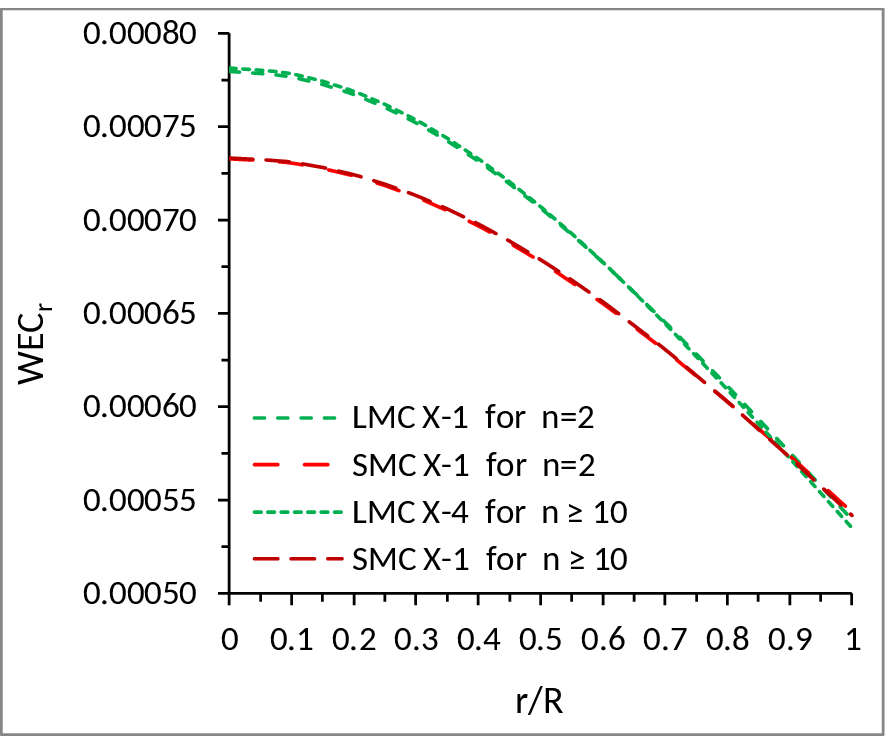} \includegraphics[width=6.5cm]{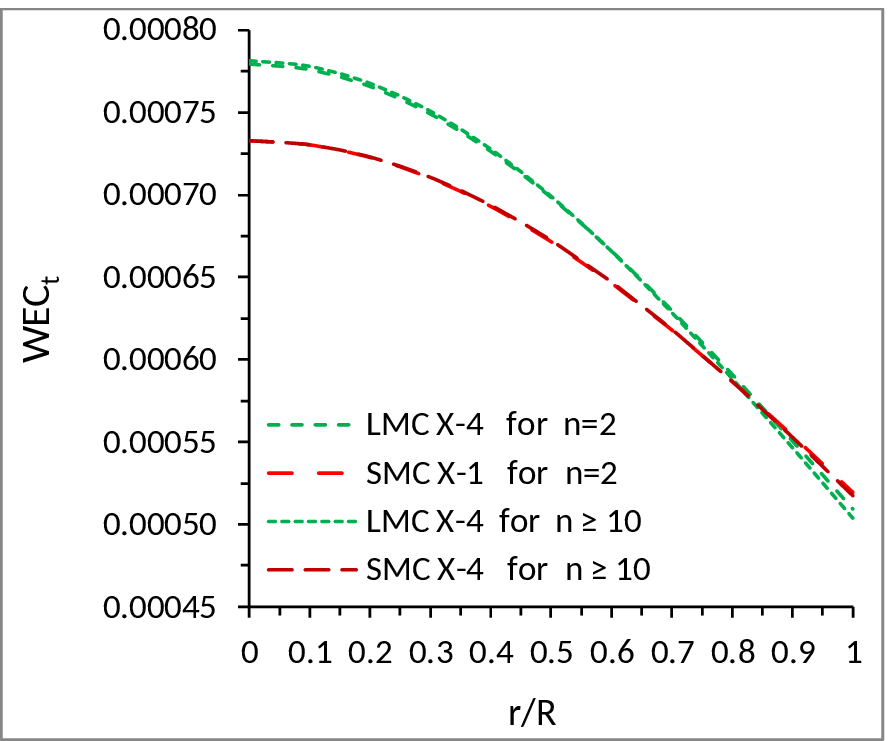}
\caption{The Behavior of energy conditions- $NEC$ (top left ), $SEC$ ( top right), $WEC_r$ (bottom left) and $WEC_t$ (bottom right) with the fractional coordinate $r/R$. }
    \label{energyconditions}
\end{figure}
%%%%%%%%%%%%%%%%%%%%%%%%%%%%%%%%%%%%%%%%%%%%%%%%%%%%%%%%%%%%%%%

\subsection{Stability of the models}

\subsubsection{Adiabatic index}
According to the condition, proposed by Heintzmann and Hillebrandt~\cite{Heintzmann1975}, for any model of anisotropic compact star the adiabatic index $\Gamma$ must be always greater than $\frac{4}{3}$ for static equilibrium. Stars are very accurately considered as adiabatic system. As there is no heat transfer from or to the compact stellar system (or any isolated gravitating object), it is considered as adiabatic system which obeys the equation $pv^\Gamma$=constant, where $\Gamma$=adiabatic index. The equation being related to pressure and volume of the system, $\Gamma$ can not have any arbitary value to maintain the stability of the system. In compact star, for a specific polytropic sequence, the mass and radius have the relationship: $R \sim M^{\frac{\Gamma-2}{3\Gamma-4}}$~\cite{Glendenning1997}. So the adiabatic index for the radial pressure can be obtained by using the following equation as
\begin{equation}
\Gamma_r=\frac{p_r+\rho}{p_r}\,\frac{dp_r}{d\rho}=\frac{p_r+\rho}{p_r}\,v^2_r. \label{eq35}
\end{equation}

Putting the values of $\rho$ and $p_r$ from Eqs. (\ref{eq18}) and (\ref{eq19}) respectively, we have
\begin{equation}
\Gamma_r=\frac{2\left[2bB(n-1)\,f^{2n+1}+a\,A\,f^{n}\,(f+2\,n\,b\,r^2)+a\,Bf(1-3br^2+4\,b\,n\,r^2)\right]\,v^2_r}
{f^{1-2n}\left(ar^2+f^{2n}\right)\,\left[4bB\,(n-1)\,f^{2n}-a\,(Bf+Af^n)\right]}. \label{eq36}
\end{equation}

Now the adiabatic index for the tangential pressure takes the following form
\begin{equation}
\Gamma_t=\frac{p_t+\rho}{p_t}\,\frac{dp_t}{d\rho}=\frac{p_t+\rho}{p_t}\,v^2_t. \label{eq37}
\end{equation}

Again plugging the values of $\rho$ and $p_t$ from Eqs. (\ref{eq18}) and (\ref{eq20}) respectively, we obtain $\Gamma_t$ as
\begin{equation}
\Gamma_t=\frac{[2f^{2n}\,(f+n\,b\,r^2)+a\,r^2\,f]\,\left[2\,b\,B\,(n-1)\,f^{2n}+a\,(B\,f+A\,f^{n})\right]\,v^2_t}
{4\,b\,B\,(n-1)f^{4n}(f+nbr^2)+a\,f^{2n}\left[-Af^{n}(f+2nb\,r^2)+B\,(b^2r^4-1)\right]}. \label{eq38}
\end{equation}

%%%%%%%%%%%%%%%%%%%%%%%%%%%%%%%%%%%%%%%%%%%%%%%%%%%%%%%%%%%%%%%
\begin{figure}[!h]\centering
    \includegraphics[width=5.5cm]{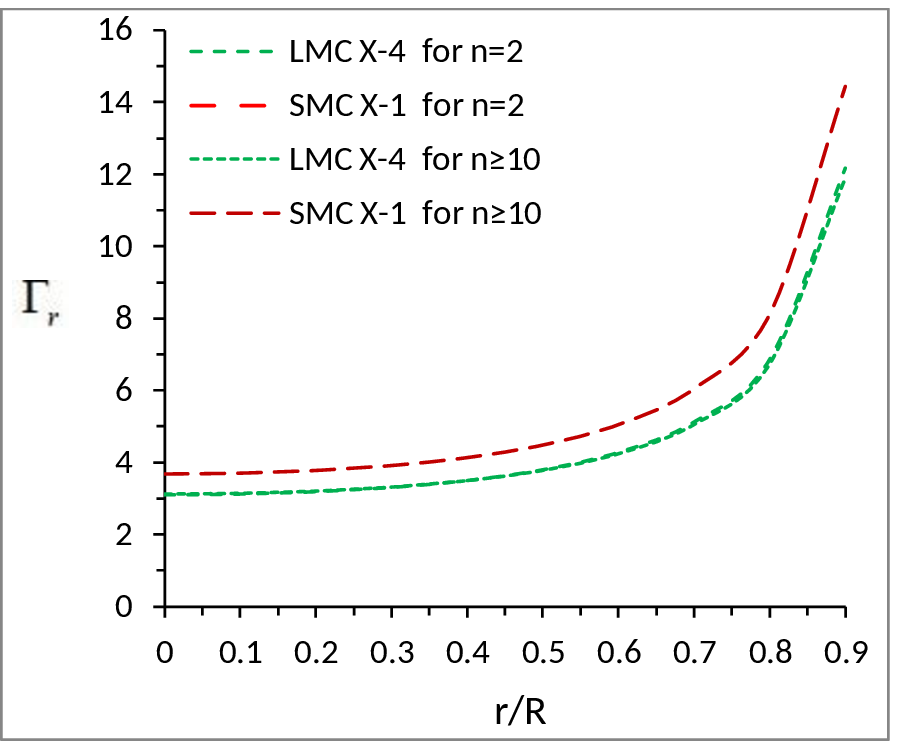} \includegraphics[width=5.5cm]{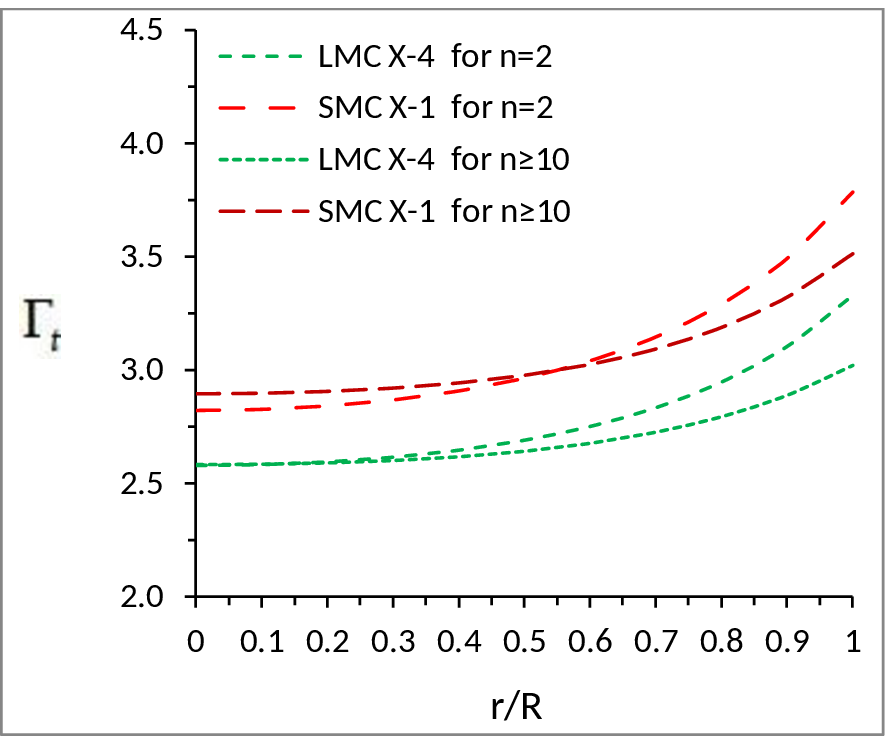}
    \caption{The Behavior of adiabatic index corresponding to radial pressure $(\Gamma_r)$ and tangential pressure $(\Gamma_t)$ with the fractional coordinate $r/R$. }
    \label{adaibaticindex}
\end{figure}
%%%%%%%%%%%%%%%%%%%%%%%%%%%%%%%%%%%%%%%%%%%%%%%%%%%%%%%%%%%%%%%

From Fig. 6, it is clear that both the radial and tangential adiabatic index follow the inequality $\Gamma_r>\Gamma_t$ throughout the star and hence our model is stable. The logic behind this is as follows: in general the anisotropy factor ($\Delta$) is a monotonic increasing function with the distance from centre to surface. Therefore the tangential pressure always dominates and increases in a higher rate than radial pressure and adiabatic index has a inverse relation with pressure so $\Gamma_r>\Gamma_t$. If we see near the surface zone ($\frac{r}{R}>0.85$), there $\Gamma_r$ increases more rapidly than $\Gamma_t$. It is because radial pressure decreases more rapidly near the surface than the tangential pressure. Also due to the boundary condition, radial adiabatic index $\Gamma_r$ will behave asymptotically very near to the surface.

\subsubsection{Herrera cracking concept}
Cracking concept was first proposed by Herrera~\cite{Herrera1992} and later on by Abereu et al.~\cite{Abreu2007} and this is a criteria for local stabilty of anisotropic compact star. Cracking is actually a local perturbation inside the anisotropic compact star due to the local anisotropy. As shown by Herrera~\cite{Herrera1992} that constant density perturbation could generate cracking on anisotropic relativistic fluids. Thus perturbation of density in some local region is not an impossible event which creates cracking and makes the system unstable. According to this cracking technique, in any local anisotropic fluid model if the radial velocity of sound is greater than the transverse velocity of sound then the region is potentially stable. For the physically acceptable system we have discussed that $V_r^2$ and $V_t^2$ must lie between 0 and 1. But from Herrera's cracking condition for the stable region the condition is $V_r^2-V_t^2 \leq1$ also to be satisfied. In our model, as shown in Fig. 7, there is no change of sign in the difference between square of sound velocity throughout the anisotropic compact star and thus confirming that the model has stable configuration.

%%%%%%%%%%%%%%%%%%%%%%%%%%%%%%%%%%%%%%%%%%%%%%%%%%%%%%%%%%%%%%%
\begin{figure}[!h]\centering
    \includegraphics[width=6.5cm]{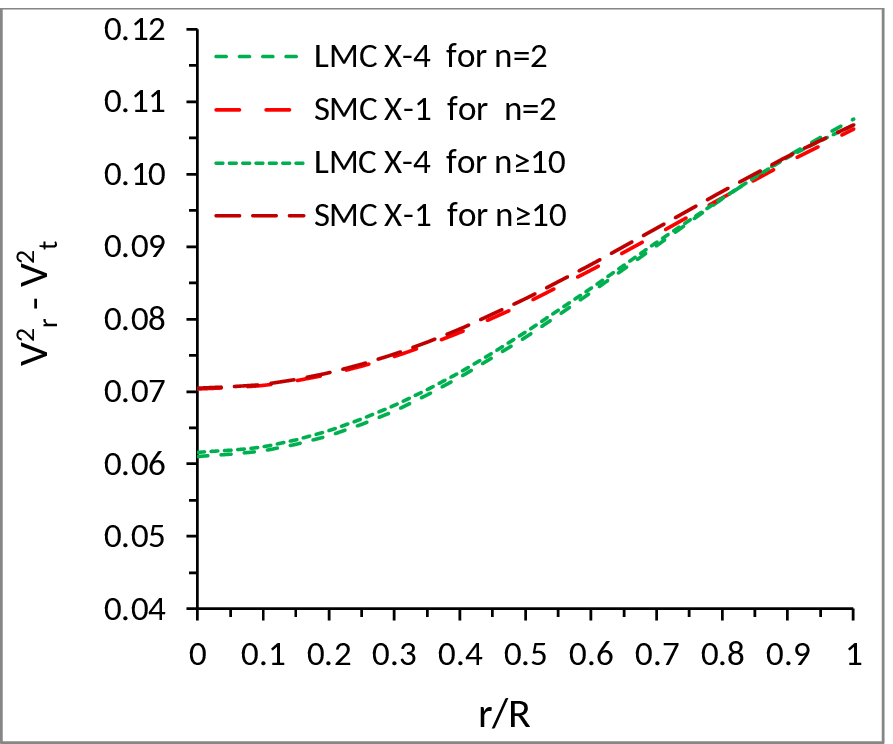} \includegraphics[width=6.5cm]{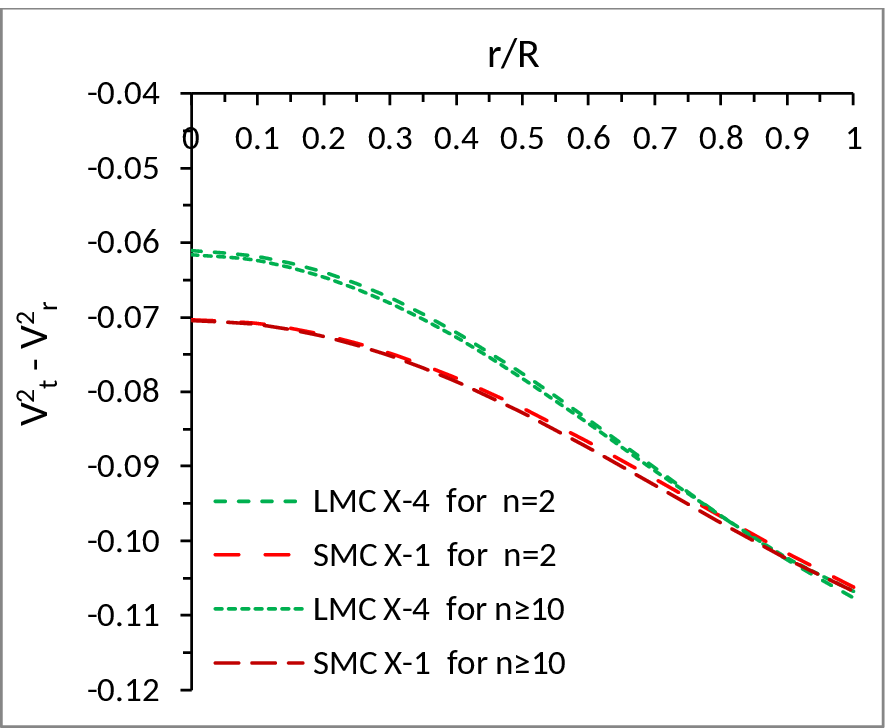}
    \caption{The variation of difference of square of velocity $V^2_r - V^2_t$ (left panel) and $V^2_t - V^2_r$ (right panel) with the fractional coordinate $r/R$. }
    \label{velocitydifference}
\end{figure}
%%%%%%%%%%%%%%%%%%%%%%%%%%%%%%%%%%%%%%%%%%%%%%%%%%%%%%%%%%%%%%%

\subsubsection{Tolman-Oppenheimer-Volkov (TOV) equation}
Anisotropy always generates a force and so there is one extra force in anisotropic compact star than comparing isotropic stellar objects. In any system, if it is stable then it is because of the force equilibrium. Therefore, for the stability of an anisotropic system the model under the three forces, viz. the gravitational force $(F_g)$, hydrostatic force $(F_h)$ and anisotropic force must satisfy the TOV equation~\cite{Tolman1939,Oppenheimer1939} which is given as
\begin{equation}
-\frac{M_G(r)(\rho+p_r)}{r}e^{\frac{\nu-\lambda}{2}}-\frac{dp_r}{dr}+\frac{2}{r}(p_t-p_r)=0, \label{eq39}
\end{equation}
where, the gravitational mass is represented by $M_G(r)$ and
can be obtained from the Einstein field equations and the Tolman-Whittaker formula as
\begin{equation}
M_G(r)=\frac{1}{2}re^{\frac{\lambda-\nu}{2}\nu'}. \label{eq40}
\end{equation}

Plugging the value of $M_G(r)$ in Eq. (\ref{eq39}), we get
\begin{equation}
-\frac{\nu'}{2}(\rho+p_r)-\frac{dp_r}{dr}+\frac{2}{r}(p_t-p_r)=0. \label{eq41}
\end{equation}

The above TOV equation therefore can be expressed into three different forces as follows $F_g+F_h+F_a=0$, where the gravitational force ($F_g$), hydrostatic force ($F_h$) and anisotropic force ($F_a$) have the forms\begin{equation}
F_g=-\frac{\nu'}{2}(\rho+p_r)=\frac{2\,b\,r\,(1-n)\,B\,(1-b\,r^2)^{-n}}{[A+B\,(1-b\,r^2)^{1-n}]}\, (p_r +\rho)\, \label{eq42}
\end{equation}

\begin{equation}
F_h=-\frac{dp_r}{dr}=-\frac{2\,r\,F_{h1} + 2\,r\,\left[b\,H(r)\,F(r) + (a- 2bn\, f^{2n-1})\,G(r)\right] \, F_{h2}}
{8\,\pi\left[B\,(1-b\,r^2) + A\,(1-b r^2)^n \right]^2 \left[a\,r^2 + (1-b r^2)^{2\,n} \right]^2}\, \label{eq43}
\end{equation}

\begin{equation}
F_a=\frac{2}{r}(p_t-p_r)=\frac{r\,\left[2bn\,f^{2n}-a\,f\,\right]\,\left[2bB\,n\,f^{2n}-aB\,f-a\,A\,f^{n}\right]}
{8\,\pi f\,\left[ar^2+f^{2n}\right]^2\,\left[B\,f+A\,f^n\right]}\, \label{eq44}
\end{equation}
with $F_{h1}=-\,b\,G(r)\, F(r)\, [8\, b\, B\, (n-1)\, n\, f^{2 n-1} + a\, H(r)]$ and $F_{h2}=[-4bB\,(n-1)\, f^{2n} + a\, G(r)]$.

In Fig. 8 nature of each of these forces has been plotted which matches the stability criteria of our anisotropic compact star model. It is shown that gravitational force is negative throughout the system due to its attractive nature and it acts towards the centre, whereas the hydrostatic and anisotropic forces are positive throughout the star beccause they are repulsive in nature, i.e. they act outwards. So the gravitational force is balanced by these two forces. Each of this forces, as obvious, is dependent on parameters $a$, $b$, $n$, $A$ and $B$ so that magnitude is different for diffrent compact star.

%%%%%%%%%%%%%%%%%%%%%%%%%%%%%%%%%%%%%%%%%%%%%%%%%%%%%%%%%%%%%%%
 \begin{figure}[!h]\centering
    \includegraphics[width=6.5cm]{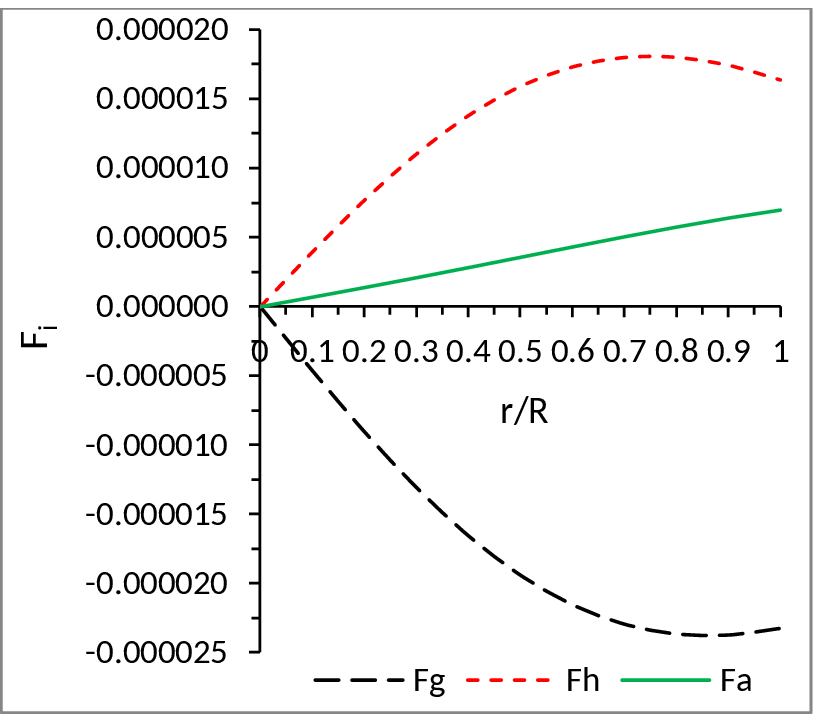} \includegraphics[width=6.5cm]{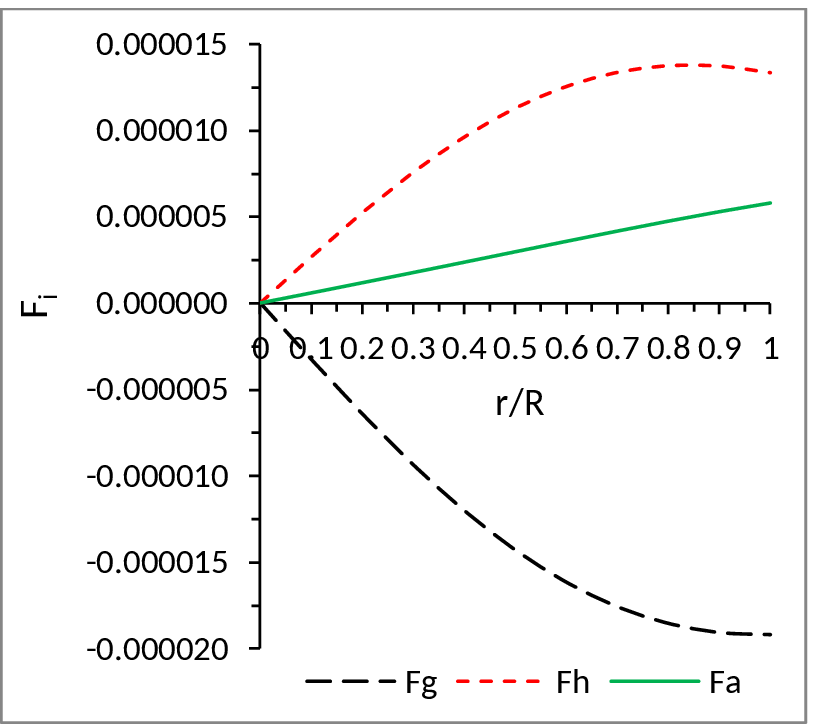}
    \includegraphics[width=6.5cm]{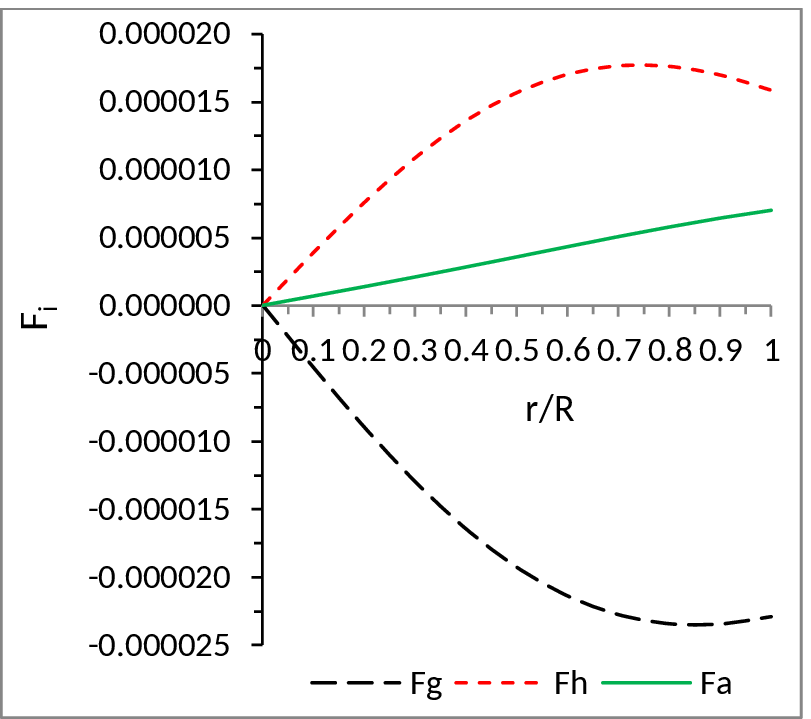} \includegraphics[width=6.5cm]{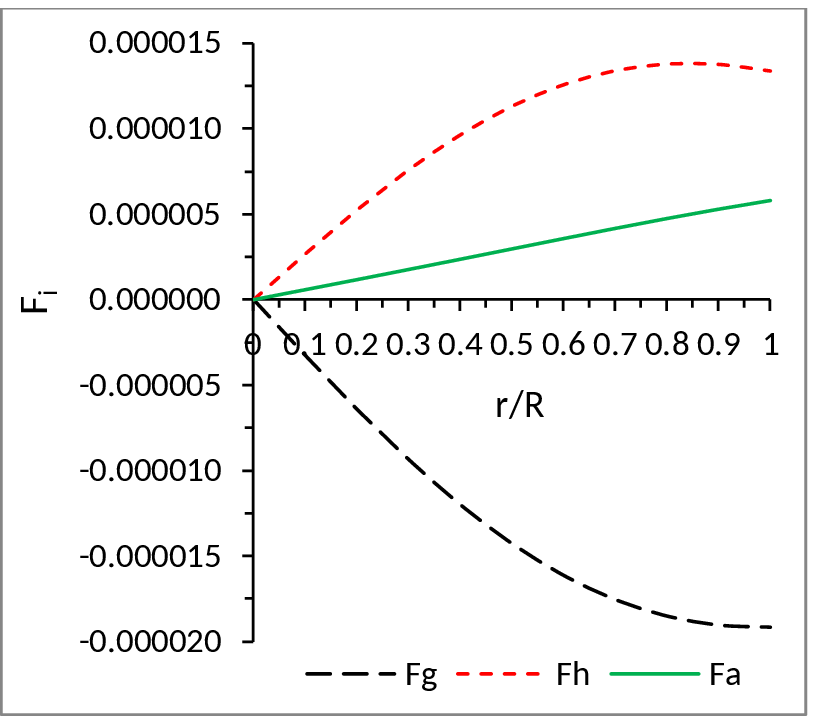}
    \caption{The variation of different forces with the fractional coordinate $r/R$. LMC X-4 for $n=2$ (top left),  SMC X-1 for n=2 (top right), LMC X-4 for $n\ge 10$ (bottom left), SMC X-1 for $n \ge 10$. }
    \label{forces}
\end{figure}
%%%%%%%%%%%%%%%%%%%%%%%%%%%%%%%%%%%%%%%%%%%%%%%%%%%%%%%%%%%%%%%

\subsection{Total mass, effective mass and compactification factor}
The metric coefficients relationship $e^{-\lambda(R)}=1-\frac{2M}{R}$ gives the expression for total mass of the anisotropic compact star as
\begin{equation}
M=\frac{\,aR^3\,}{2\,[aR^2+(1-bR^2)^{2n}]}.    \label{eq26}
\end{equation}

On the other hand, the effective mass of the anisotropic star can be calculated as follows
\begin{equation}
m_{eff}=4\pi{\int^R_0{\rho\,r^2\,dr}}=\frac{aR^3}{2\,[aR^2+(1-bR^2)^{2n}]}. \label{eq45}
\end{equation}

By using the above formula the compactification factor $u(r)$ can be obtained as
\begin{equation}
u(r)=\frac{m_{eff}(r)}{r}=\frac{ar^2}{2\,[ar^2+(1-br^2)^{2n}]}. \label{eq46}
\end{equation}

So larger the radius of the compact star the smaller will be the value of the metric coefficient $e^{-\lambda}$ or larger will be the value of the metric coefficient $e^\nu$. Thus smaller the value of the metric coefficient $e^\nu$, more will be the compactness of the compact star. 

The expression in Eq. (\ref{eq26}) actually defines the compactness parameter of a compact star, i.e. larger the value of the $\frac{M}{R},$ more will be the compactness of the star. The parameters $a$ and $b$ both play a significant role for the mass and hence compactness of a compact star. For a fixed assumed value of $a$, larger the value of the parameter $b$, lower will be the mass of a compact star as the denominator conatains $(1-b r^2)^{2n}$ as can be seen from  Eq. (\ref{eq46}). So it will be always positive irrespective of the value of $b$. Hence we observe that the parameter $b$ plays a negative role in the mass of a compact star and simillarly the index $n$ which makes decrease of the mass of a compac star. In contrast to the increase of the value of the other parameter $a$, the mass of the compact star increases.

\subsection{Surface redshift}
In astrophysics redshift is a very important phenomenon as it is used several times for detecting objects which are far away from us. Centre is farthest position from observer and hence redshift is maximum for the centre whereas it is minimum near the surface. Compactness and redshift are related with each other as the expression of redshift is also the ratio of mass and radius. So higher the redshift means higher is the compactness. The more elaborated relationship of redshift is given below
\begin{equation}
Z(r)= \sqrt{[A+B\,(1-br^2)^{1-n}]^{-1}}-1. \label{eq47}
\end{equation}

The above Eq. (\ref{eq47}) says that redshift $Z$ is dependent on the parameter $b$, constants $A$ as well as $B$ and index $n$. So for different anisotropic compact star, depending on these factors, redshift will be different. According to the work of Ivanov~\cite{Ivanov2002}, surface redshift can not exceed the value 3.842 for tangential pressure to satisfy SEC. Surface redshift and $\frac{M}{R}$ exceed the corresponding limit for perfect fluid as usually happens for anisotropic sphere.

The variation of the surface redshift with the fractional radial parameter are shown in Fig. 9. This figure shows that the surface redshift is monotonically decreasing in nature. For higher value of $n$, the redshift is large. The maximum value of the redshift in our model is below that magnitude claimed by various authers for which it is believed to provide stability.

%%%%%%%%%%%%%%%%%%%%%%%%%%%%%%%%%%%%%%%%%%%%%%%%%%%%%%%%%%%%%%%
\begin{figure}[!h]\centering
    \includegraphics[width=6.5cm]{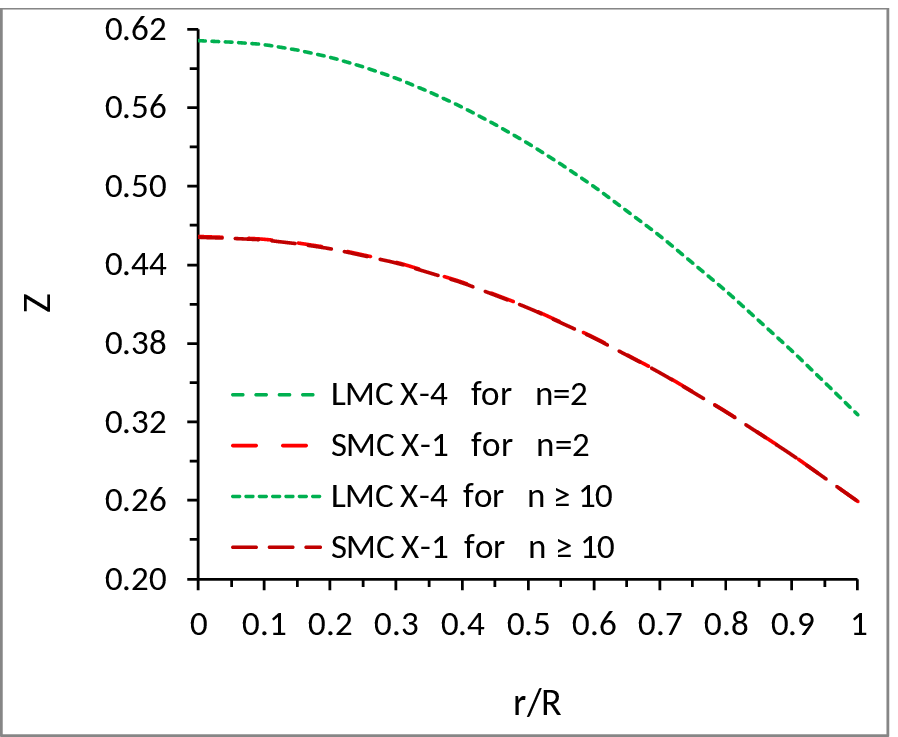}
    \caption{The variation of redshift ($Z$) with the fractional coordinate $r/R$. }
    \label{redshift}
\end{figure}
%%%%%%%%%%%%%%%%%%%%%%%%%%%%%%%%%%%%%%%%%%%%%%%%%%%%%%%%%%%%%%%

\begin{table}
\centering
\caption{Values of parameters or constants for (i) LMC X-4 with mass $M=1.2928\pm0.00033 M_{\odot}$ and radius $R=8.831 Km$, (ii) SMC X-1  with mass $M=1.04\pm 0.002 M_{\odot}$ and radius $R=8.301 Km$}  \label{Table 1}
{\begin{tabular}{@{}c|cccc|cccc@{}}
\hline & LMC X-4 & &   &  & SMC X-1 \\
\hline n &$a(km^{-2})$&$b (km^{-2})$&$A$&$B$&$a(km^{-2})$ & $b(km^{-2})$ & $A$ &$B$\\
\hline 2 & 7.5$\times{10}^{-3}$ & 8.21$\times{10}^{-4}$ & -1.3470 & 1.9639 & 6.81$\times{10}^{-3}$ & 7.8$\times{10}^{-4}$&  1.2518 & 1.9361 \\
\hline 10 & 7.5$\times{10}^{-3}$ & 1.64$\times{10}^{-4}$ & -0.4695 & 1.0901 & 6.81$\times{10}^{-3}$ &1.59$\times{10}^{-4}$& -0.3707 & 1.0551 \\
\hline 100 & 7.5$\times{10}^{-3}$ & 1.64$\times{10}^{-5}$&  -0.3697  & 0.9906 & 6.81$\times{10}^{-3}$ &1.59$\times{10}^{-5}$&-0.2743 & 0.9588 \\
\hline 1000 & 7.5$\times{10}^{-3}$ & 1.64$\times{10}^{-6}$ & -0.3607 &  0.9816 & 6.81$\times{10}^{-3}$ &1.59$\times{10}^{-6}$& -0.2656 & 0.9502 \\
\hline 10000 & 7.5$\times{10}^{-3}$ & 1.64$\times{10}^{-7}$ & -0.3598 & 0.9808 & 6.81$\times{10}^{-3}$ &1.59$\times{10}^{-7}$&-0.2647 & 0.9493 \\
\hline 20000 & 7.5$\times{10}^{-3}$ & 8.2$\times{10}^{-8}$ &  -0.3597 & 0.9807 & 6.81$\times{10}^{-3}$ &7.95$\times{10}^{-8}$&-0.2647 & 0.9493 \\
\hline
\end{tabular}}
\end{table}
%%%%%%%%%%%%%%%%%%%%%%%%%%%%%%%%%%%%%%%%%%%%%%%%%%%%%%%%%%%%%%%

%%%%%%%%%%%%%%%%%%%%%%%%%%%%%%%%%%%%%%%%%%%%%%%%%%%%%%%%%%%%%%%
\begin{table}
\centering \caption{The Matter densities and central pressure for
different  compact star candidates for different values of $n$}
\label{Table4}
{\begin{tabular}{@{}c|c|cccccc@{}}
\hline
$n$ & Compact star & Central density & Surface density & Central pressure \\
   & & ($gm/cm^{3}$) & ($gm/cm^{3}$) & ($dyne/cm^{2}$) \\ \hline

$n=2$ & LMC X-4 & 1.2083$\times{10}^{15}$ & 7.2812$\times{10}^{14}$ & 1.4101$\times{10}^{35}$ \\

 & SMC X-1 & 1.0971$\times{10}^{15}$  & 7.3319$\times{10}^{14}$ & 9.7545$\times{10}^{35}$ \\ \hline

 $n\ge10$ & LMC X-4 & 1.2083$\times{10}^{15}$ & 7.2177$\times{10}^{14}$ & 1.3877$\times{10}^{35}$  \\

 & SMC X-1 & 1.0971$\times{10}^{15}$  & 7.3095$\times{10}^{14}$ & 9.7401$\times{10}^{35}$ \\ \hline
\end{tabular}}
\end{table}

%%%%%%%%%%%%%%%%%%%%%%%%%%%%%%%%%%%%%%%%%%%%%%%%%%%%%%%%%%%%%%

\section{Conclusion}
In the present paper we have performed a study on the compact star with anisotropy and class 1 metric under the framework of Einstein's theory of general relativity. This yields a set of solutions which are interesting and physically viable as far as properties of the compact stars are concerned. The investigations on the physical features of the model include the issues like (i) regularity behavior of stars at the centre, (ii) well behaved condition for velocity of sound, (iii) energy conditions, (iv) stabilty of the system via the following three techniques - adiabatic index, Herrera cracking concept and TOV equation, (v) total mass, effective mass and compactification factor and (vi) surface redshift. We would like to highlight some of the salient features of our model below.\\

 {\bf 1. Regularity behavior of stars at the centre}:\\
{\bf(i)} By putting $r=0$ in Eqs. (\ref{eq16}) and (\ref{eq17}), we can get $e^{\nu(0)}=(A+B)^2$ and $e^{\lambda(0)}=1$ which are finite at the centre and hence free from singularity. 

{\bf(ii)} For any physically acceptable model, $p_r$ and $p_t$ must be positive and finite at the centre which for the present model provides the condition $\frac{A}{B} < \frac{4\,b\,(n-1)-a}{a}$.

{\bf(iii)} From Eq. (18) matter density at centre is  $\rho(0)=\frac{3\,a}{8\,\pi}>0$, that implies $a$ must be positive and hence the central pressure is also positive and finite. 

{\bf(iv)} The solution of anisotropic compact star must satisfy the Zel'dovich condition, i.e.  $p_r/\rho <1$ and $p_t/\rho < 1$. Using the Eqs. (\ref{eq27}) and (\ref{eq28}), we get $\frac{b\,(n-1)-a}{a} < \frac{A}{B} < \frac{4\,b\,(n-1)-a}{a}$, where $n>1$.\\

{\bf 2. Well behaved condition for velocity of sound}:\\
In the present anisotropic compact stellar model both the radial and tangential sound velocity are maximum at the centre and monotonically decrease upto the surface. Now the causality conditions demands that the radial and transverse velocity of sound must be less than unity. It can be observed from Fig. 4 that the values of $V_r$ and  $V_t$ are lies between 0 and 1 everywhere inside the stellar system. \\

{\bf 3. Energy conditions}:\\
For any physically viable relativistic object the energy conditions, viz. Null Energy Condition (NEC), Strong Energy Condition (SEC) and Weak Energy Condition (WEC) are to be satisfied by the matter inside it. From Fig. 5  it can be observed that energy conditions are satisfied in the present model.\\

{\bf 4. Stabilty of the system via the following three techniques}:

{\bf (i) Adiabatic index}:
According to Heintzmann and Hillebrandt~\cite{Heintzmann1975}, for any model of anisotropic compact star the adiabatic index $\Gamma$ must be always greater than $\frac{4}{3}$ for static equilibrium. We have calculated $\Gamma_r$ and $\Gamma_t$ which are plotted in Fig. 6. From this figure it is clear that both the radial and tangential adiabatic index follow the inequality $\Gamma_r>\Gamma_t$ throughout the star and hence our model is stable. 

{\bf (ii) Herrera cracking concept}:
According to this cracking technique if the radial velocity of sound is greater than the transverse velocity of sound then the region is potentially stable and $V_r^2$ and $V_t^2$ must lie between 0 and 1. But from Herrera's cracking condition for the stable region the condition is $V_r^2-V_t^2 \leq1$ also to be satisfied. From Fig. 7 we note that there is no change of sign in the difference between square of sound velocity throughout the anisotropic compact star and thus confirming that the model has stable configuration.

{\bf (iii) TOV equation}:
The TOV equation which can be expressed into three different forces as follows $F_g+F_h+F_a=0$, where the gravitational force ($F_g$), hydrostatic force ($F_h$) and anisotropic force ($F_a$) are plotted in Fig. 8. It is shown that gravitational force is negative throughout the system due to its attractive nature and it acts towards the centre. On the other hand, the hydrostatic and anisotropic forces are positive throughout the sphere due to repulsive nature, i.e. they act outwards. \\

{\bf 5. Total mass, effective mass and compactification factor}:\\
We have found out expression for the total mass, effective mass and compactification factor of the compact star from our propsed model. It is observed that each of these physical quantities are acceptable in their physical realm. It is also noted that larger the radius of the compact star the smaller will be the value of the metric coefficient $e^{-\lambda}$ or larger will be the value of the metric coefficient $e^\nu$. This indicates that smaller the value of the metric coefficient $e^\nu$, more will be the compactness of the compact star. \\

{\bf 6. Surface redshift}:\\
As the redshift $Z$ is dependent on the parameter $b$, constants $A$ as well as $B$ and index $n$, so for different anisotropic compact star, depending on these factors, redshift will be different. Fig. 9 shows that the surface redshift is monotonically decreasing in nature.

Based on above discussions we may conclude that our presented model for anisotropic compact stars are physically viable and hence acceptable. The Tables 1 and 2 also support this statement in connection to the results obtained therein for different physical parameters. 

\section*{Acknowledgments}
SKM acknowledges support from the authority of University of Nizwa, Nizwa, Sultanate of Oman. SR is thankful to the Inter-University Centre
for Astronomy and Astrophysics (IUCAA), Pune, India for providing Visiting Associateship under which preliminary part of this work was carried out. SR also expresses his gratitude to the authority of University of Nizwa for providing all types of working facilities and hospitality under a short term visit where the final stage of the work has been performed.


\begin{thebibliography}{99}

\bibitem{Rago1991} H. Rago, Astrophys. Space Sci. {\bf 183}, 333 (1991)

\bibitem{BL1974} R.L. Bowers, E.P.T. Liang, Class. Astrophys. J. {\bf 188}, 657 (1974)

\bibitem{Ruderman1972} R. Ruderman, Rev. Astron. Astrophys. \textbf{10}, 427 (1972)

\bibitem{Canuto1973} V. Canuto, Neutron Stars: General Review Solvay Conf. on Astrophysics
and Gravitation (Brussels, Sept., 1973)

\bibitem{HS1997} L. Herrera, N.O. Santos, Phys. Report. \textbf{286}, 53 (1997)

\bibitem{Ivanov2002} B.V. Ivanov, Phys. Rev. D \textbf{65}, 104011 (2002)

\bibitem{SM2003} F.E. Schunck, E.W. Mielke, Class. Quantum Gravit. \textbf{20}, 301 (2003)

\bibitem{MH2003} M.K. Mak, T. Harko, Proc. R. Soc. A \textbf{459}, 393 (2003)

\bibitem{Usov2004} V.V. Usov, Phys. Rev. D \textbf{70}, 067301 (2004)

\bibitem{Varela2010} V. Varela, F. Rahaman, S. Ray, K. Chakraborty, M. Kalam, Phys. Rev. D \textbf{82}, 044052
(2010)

\bibitem{Rahaman2010} F. Rahaman, S. Ray, A.K. Jafry, K. Chakraborty, Phys. Rev. D \textbf{82}, 104055 (2010)

\bibitem{Rahaman2011} F. Rahaman, P.K.F. Kuhfittig, M. Kalam, A.A. Usmani, S. Ray, Class. Quantum Gravit.
\textbf{28}, 155021 (2011)

\bibitem{Rahaman2012} F. Rahaman, R. Maulick , A.K. Yadav, S. Ray, R. Sharma, Gen. Relativ. Gravit. \textbf{44}, 107
(2012)

\bibitem{Kalam2012} M. Kalam, F. Rahaman, S. Ray, Sk.M. Hossein, I. Karar, J. Naskar, Eur. Phys. J.
C \textbf{72}, 2248 (2012)

\bibitem{Deb2015} D. Deb, S.R. Chowdhury, S. Ray, F. Rahaman, arXiv:1509.00401 [gr-qc]
	
\bibitem{Shee2016} D. Shee, F. Rahaman, B.K. Guha, S. Ray, Astrophys. Space Sci. \textbf{361}, 167 (2016)

\bibitem{Maurya2016} S.K. Maurya, Y.K. Gupta, S. Ray, D. Deb, Eur. Phys. J. C \textbf{76}, 693 (2016)

\bibitem{Maurya2017} S.K. Maurya, D. Deb, S. Ray, P.K.F. Kuhfittig, arXiv:1703.08436

\bibitem{Eddigton1924} A.S. Eddington, The Mathematical Theory of Relativity (Cambridge University Press,
Cambridge, 1924)

\bibitem{Karmarkar1948} K.R. Karmarkar, Proc. Indian Acad. Sci. A {\bf 27}, 56 (1948)

\bibitem{Schlaefli1871} L. Schl{\"a}fli, Nota alla memoria del. Sig. Beltrami, sugli spazii di curvatura constante, Ann. di mat., second series, \textbf{5}, 170 (1871–1873)

\bibitem{Kuzeev1980} R.R. Kuzeev, Gravit. Teor. Otnosit. 16, 93 (1980)

\bibitem{SKM1} S.K. Maurya, Y.K. Gupta, S. Ray, D. Deb, Euro. Phys. J. C, {\bf 76}, 693 (2016)

\bibitem{SKM2} S.K. Maurya, Y.K. Gupta, S. Ray, D. Deb, Euro. Phys. J. C, {\bf 77}, 45 (2017)

\bibitem{Wheeler1998} J.A. Wheeler, Geons, Black Holes and Quantum Foam (New Yok: W.W. Norton, p. 235, 1998) 

\bibitem{SP2016} K.N. Singh, N. Pant, Euro. Phys. J. C, {\bf 76}, 524 (2016)

\bibitem{Misner1964} C.W. Misner, D.H. Sharp, Phys. Rev. B \textbf{136}, 571 (1964)

\bibitem{Heintzmann1975} H. Heintzmann, W. Hillebrandt, Astron. Astrophys. \textbf{38}, 51 (1975)

\bibitem{Glendenning1997} N.K. Glendenning, Compact Stars: Nuclear Physics, Particle Physics and General Relativity (Springer, New York, 1997)

\bibitem{Herrera1992} L. Herrera, Phys. Lett. A \textbf{165}, 206 (1992)

\bibitem{Abreu2007} H. Abreu, H. Hernandez, L.A. Nunez, Class. Quantum Gravit. \textbf{24}, 4631 (2007)

\bibitem{Tolman1939} R.C. Tolman, Phys. Rev. {\bf 55}, 364 (1939)

\bibitem{Oppenheimer1939} J.R. Oppenheimer, G.M. Volkoff, Phys. Rev. {\bf 55}, 374 (1939)

\end{thebibliography}
\end{document}